\documentclass[preprint2]{emulateapj}

\newcommand{\myemail}{enrique.lopezrodriguez@utsa.edu}
\newcommand{\degree}{$^{\circ}$}
\newcommand{\um}{$\mu$m}

\slugcomment{Accepted for publication in ApJ}

\shorttitle{Polarized Mid-Infrared Synchrotron Emission in the Core of Cygnus A}
\shortauthors{Lopez-Rodriguez et al.}

\begin{document}


\title{Polarized mid-infrared synchrotron emission in the core of Cygnus A}


\author{E. Lopez-Rodriguez\altaffilmark{1}\altaffilmark{$\star$}, C. Packham\altaffilmark{1}, C. Tadhunter\altaffilmark{2}, R. Mason\altaffilmark{3}, E. Perlman\altaffilmark{4}, A. Alonso-Herrero\altaffilmark{5}\altaffilmark{$\dagger$}, C. Ramos Almeida\altaffilmark{6,7,$\dagger\dagger$}, K. Ichikawa\altaffilmark{8}, N. A. Levenson\altaffilmark{9}, J.~M. Rodr\'iguez-Espinosa\altaffilmark{6,7}, C. A. \'Alvarez\altaffilmark{7,10}, E. A. Ram\'irez\altaffilmark{11}, C.~M. Telesco\altaffilmark{12}}

\affil{$^{1}$Department of Physics \& Astronomy, University of Texas at San Antonio, One UTSA Circle, San Antonio, TX 78249, USA \\
	$^{2}$Department of Physics \& Astronomy, University of Sheffield, Sheffield S3 7RH \\
	$^{3}$Gemini Observatory, Northern Operations Center, 670 N. A'ohoku Place, Hilo, HI 96720, USA \\
	$^{4}$Department of Physics and Space Sciences, Florida Institute of Technology, Melbourne, FL 32901 \\
	$^{5}$Instituto de F\'isica de Cantabria, CSIC-UC, 39005 Cantabria, Spain \\
	$^{6}$Instituto de Astrof\'isica de Canarias, Calle V\'ia L\'actea s/n, 38205, Tenerife, Spain \\
	$^{7}$Universidad de La Laguna, Departamento de Astrof\'isica, E-38206 La Laguna, Tenerife, Spain \\
	$^{8}$Department of Astronomy, Graduate School of Science, Kyoto University, Kitashirakawa-Oiwake cho, Kyoto 606-8502, Japan \\
	$^{9}$Gemini Observatory, Casilla 603, La Serena, Chile \\
	$^{10}$GTC Project, Instituto de Astrof\'isica de Canarias (IAC), 38200 La Laguna, Tenerife, Spain. \\
	$^{11}$Universidade de S\~ao Paulo, IAG, Rua do Mat\~ao 1226, Cidade Universit\'aria, S\~ao Paulo 05508-900, Brazil. \\
	$^{12}$Department of Astronomy, University of Florida, 211 Bryant Space Science Center, P.O. Box 11205, Gainesville, FL 32611-2055, USA
}


\altaffiltext{$\star$}{Email: \myemail}
\altaffiltext{$\dagger$}{August G. Linares Senior Research Fellow}
\altaffiltext{$\dagger\dagger$}{Marie Curie Fellow}


\begin{abstract}

We present high-angular ($\sim$0.4$''$) resolution mid-infrared (MIR) polarimetric observations in the 8.7 \um~and 11.6 \um~filters of Cygnus A using CanariCam on the 10.4-m {\it Gran Telescopio CANARIAS}. A highly polarized nucleus is observed with a degree of polarization of $11\pm3\%$ and $12\pm3\%$ and position angle of polarization of $27\pm8^{\circ}$~and $35\pm8^{\circ}$~in a 0.38$''$ ($\sim$380 pc) aperture for each filter. The observed rising of the polarized flux density with increasing wavelength is consistent with synchrotron radiation from the pc-scale jet close to the core of Cygnus A. Based on our polarization model, the synchrotron emission from the pc-scale jet is estimated to be 14\% and 17\% of the total flux density in the 8.7 \um~and 11.6 \um~filters, respectively. A blackbody component with a characteristic temperature of 220 K accounts for $>$75\% of the observed MIR total flux density. The blackbody emission arises from a combination of (1) dust emission in the torus; and (2) diffuse dust emission around the nuclear region, but the contributions of the two components cannot be well constrained in these observations.

\end{abstract}






\section{Introduction}

Little is known about the mid-infrared (MIR) polarization at high-angular resolution of Active Galactic Nuclei (AGN). The only high-angular resolution MIR polarimetric observations of an AGN were published by \citet{Packham2007}, who used the polarimetric mode of MICHELLE with the  9.7 \um~filter on the 8.1-m {\it Gemini North} telescope. This study revealed complex polarization structures in the inner 2$''$ of NGC 1068. However, only one filter was used, making a full interpretation of the different mechanisms of polarization difficult. The polarization mechanisms and hence the physical mechanisms and the nature of the inner region of the AGN can be disentangled only through multi-wavelength MIR polarimetric observations \citep{Aitken2004}.

At a redshift of 0.0562 (\citet{SRL1994}, $H_{0} = 73$ km s$^{-1}$ Mpc$^{-1}$; 1$''$ $\sim$ 1 kpc), Cygnus A is one of the most studied Faranoff-Riley class II (FRII) radio galaxies \citep{CB1997}.  Cygnus A shows complex structures: including a core with a patchy dust lane, ionization cones, and jets. A compact, unresolved nucleus, with a peak in the luminosity at IR wavelengths was observed using the Palomar 5.08-m Hale Telescope \citep{Djorgovski1991}. Based on their reported visual extinction A$_{V} =$ 50$\pm$30 mag and the IR luminosity, Cygnus A should be classified as quasar. The classification as a quasar has also been made by several others  \citep[e.g.][]{Ueno1994,Tadhunter1999}, with different estimations of the visual extinction to the nucleus of Cygnus A. For instance, \citet{Ueno1994} obtained a visual extinction of 170$\pm$30 mag from modeling the X-ray spectrum. \citet{Simpson1995} compared the [OIII] emission lines, MIR continuum and hard X-ray continuum and estimated a visual extinction of 143$\pm$35 mag. Using high-spatial resolution {\it HST}/NICMOS observations in the 2.0 \um~filter and the near-infrared (NIR) to X-ray correlation \citep{Kriss1988} for quasars, \citet{Tadhunter1999} estimated a visual extinction of 94 mag. The discrepancy between visual extinctions at different wavelengths can be accounted for as different wavelengths penetrate through different depths and structures in and around the central engine. The X-ray values represent the best estimate for the total extinction to the central engine of Cygnus A, while NIR values are dominated by emission line and/or hot (T$\sim$1000 K) dust emission close to the central engine. The MIR emission dominantly arises from warm (T$\sim$300 K) dust within the torus and/or dust emission from extended dust component. 

Although optical total flux emission (as opposed to polarized flux emission) from the central engine is not observed, optical polarized broad emission lines have been observed \citep[e.g.][]{,Tadhunter1990,JT1993,Ogle1997}. This is most readily interpreted by the presence of broad emission lines in the central engine scattered into our line of sight (LOS), which would otherwise be obscured by the geometrically and optically thick and dusty torus. Such an interpretation is entirely consistent with the  unified model of AGN \citep{Antonucci1993,UP1995}. Also, it shows the potential of polarimetric techniques to investigate the core of Cygnus A. Previous studies \citep{Tadhunter1990} observed, in the V-band, a centrosymmetric polarization pattern along the ionization cones in the inner $\sim$3.5$''$ of Cygnus A. This polarization pattern is the signature of a central point source core whose radiation is polarized through scattering by dust and/or electrons. \citet{JT1993} observed an optical (0.48-0.78 \um) polarized spectrum rising to the blue, suggesting that dust scattering is the dominant mechanism of polarization responsible for the extended optical polarized flux of Cygnus A. Through spectropolarimetric observations in the range of 0.3-0.85 \um, \citet{Ogle1997} investigated the broad and narrow emission lines and the color and high polarization of the scattered continuum. They concluded that all features are consistent with scattering by dust. Thus, an extended dusty component is dominantly responsible for the polarized flux through dust scattering at optical wavelengths.

Further high-angular resolution {\it HST}/NICMOS observations using the 1.1 \um, 1.6 \um, 2.0 \um~and 2.25 \um~filters detailed the region of the AGN and the ionization cones \citep{Tadhunter1999}. This study detected a X-shape structure consistent with straight lines in the central 2$''$ of Cygnus A. This structure was also observed in 10.8 \um~ and 18.2 \um~broad-filters imaging observations using OSCIR on the 10.0-m {\it Keck II} telescope \citep{Radomski2002}, as well as in 11.7 \um~narrow-filter imaging observations using the Long Wavelength Spectrometer (LWS) on the 10.0-m {\it Keck I} telescope \citep{WA2004}. Imaging polarimetric observations in the 2.0 \um~filter using {\it HST}/NICMOS detected highly polarized extended emission with a maximum in the degree of polarization of $\sim$25\%,  spatially coincident with one arm of the biconical structures seen in total flux \citep[][hereafter T00]{Tadhunter2000}. They suggested that scattering is the dominant polarizing mechanism in the 2.0 \um~filter and that the detection of only one arm in polarized flux could be produced by intrinsic anisotropy from the central engine, possibly a warped disk. 

Several studies have investigated the different components within the unresolved nucleus of Cygnus A using polarimetric techniques. In the 2.0 \um~filter, T00 measured the nucleus to have a degree of polarization of $\sim$20\% at the flux peak and $\sim$10\% in a 0.375$''$ $\times$ 0.375$''$ aperture. This is most readily accounted for by an unresolved scattering region close to the central engine. However, as only a single wavelength was available, definitive conclusions about the polarization mechanisms could not be drawn. A kinematic study \citep{vanBemmel2003} of the polarized [OIII] emission lines showed that a dust scattering component arises from spatially unresolved optically thin dusty clouds moving away from the nucleus of Cygnus A, and along the dusty biconical structures observed at NIR \citep{Tadhunter1999} and MIR \citep{Radomski2002,WA2004} wavelengths. Although radio polarimetric observations of Cygnus A have not been published, pc-scale jets have been observed \citep{Krichbaum1998} through VLBI radio observations in Cygnus A, consistent with the classification as a FRII source and with the presence of a synchrotron component close to the AGN.

In total flux, recent studies \citep{Privon2012,Merlo2014} have modeled the spectral energy distribution (SED) of the central regions of Cygnus A. \citet[][hereafter P12]{Privon2012} suggested that the SED can be explained by 1) star formation, dominant in the far-IR (FIR); 2) clumpy torus emission with a diameter of $\sim$130 pc, dominant in the MIR; and 3) a cut-off wavelength synchrotron component with a break in the MIR. Through high-angular resolution N-band spectroscopic observations using the 8.2-m {\it Subaru} telescope, \citet[][]{Merlo2014} suggested that the MIR total flux emission can be explained by a blackbody component of 217$\pm$3 K located in our LOS to the AGN. To investigate the MIR core of Cygnus A and to examine the strength of the synchrotron jet further MIR polarimetry can be a powerful tool.

In this paper, we aim to explain the dominant mechanism of polarization of Cygnus A at MIR wavelengths. We present high-angular ($\sim$0.4$''$) resolution polarimetric observations in the 8.7 \um~and 11.6 \um~filters using CanariCam on the 10.4-m {\it Gran Telescopio CANARIAS} ({\it GTC}). This paper is organized as follows. We describe the observations and data reduction in $\S$2 and the results are presented in $\S$3. A polarization model to account for the IR polarization of Cygnus A is presented in $\S$4. $\S$5 presents the discussion of our results and $\S$6 presents the conclusions.



\section{Observations and data reduction}
\label{OBS}

Cygnus A was observed on 2012 August 12 during commissioning of the imaging polarimetric mode \citep{Packham2005b} of CanariCam \citep{Telesco2003} on the 10.4-m {\it GTC} in La Palma, Spain. CanariCam uses a Raytheon 320 $\times$ 240 pixels Si:As array, with a pixel scale of 0.0798$''$ pixel$^{-1}$. The polarimetric mode of CanariCam uses a half-wave plate (HWP), a field mask, and a Wollaston prism. The Wollaston prism and HWP are made from sulphur-free CdSe. The HWP has a chromatic dependency on the polarization retardation, resulting in a variable polarimetric efficiency across the wavelength range of 7.5 - 13 \um.  However, this has been well determined\footnote{Further information about the polarization efficiency is at: http://www.gtc.iac.es/instruments/canaricam/~\#Polarization\_Measurement\_Efficiency}. In standard polarimetric observations, the HWP is rotated in 4 position angles (PA) in the following sequence: 0\degree, 45\degree, 22.5\degree, and 67.5\degree. The field mask consists of a series of slots of 320 $\times$ 25 pixels each, corresponding to a field of view (FOV) of 25.6$''$~$\times$~2.0$''$, where a total of three slots can be used, providing a non-contiguous total FOV of 25.6$''$~$\times$~6.0$''$. 

The Si2 ($\lambda_{c}=$ 8.7 \um, $\Delta\lambda=$ 1.1 \um, 50\% cut-on/off) and Si5  ($\lambda_{c}=$ 11.6 \um, $\Delta\lambda=$ 0.9 \um, 50\% cut-on/off) filters provide the best combination of sensitivity, spatial resolution and spread in wavelength within the instrument filter set, so these filters were used for the observations. Observations were made using a standard chop-nod technique to remove time-variable sky background and telescope thermal emission, and to reduce the effect of 1/{\it f} noise from the array. In all observations, the chop-throw was 8$''$, the chop-angle was 105\degree~E of N to locate the extended dust emission of Cygnus A along the FOV of the slot and the chop frequency was 1.93 Hz. The angle of the short axis of the array with respect to the North on the sky (i.e. instrumental position angle) was 15\degree~E of N and the telescope was nodded every 45.5s along the chopping direction. Only one slot with a FOV of 25.6$''$~$\times$~2.0$''$ was used in these observations. In both filters, we took two sets of images with an on-source time of 582s each. To improve the signal-to-noise ratio (SNR) in the final image, the negative images (produced by the chop-nod technique) on the array were also used, providing an useful on-source time of 2328s for each filter. 

Data were reduced using custom IDL routines. The difference for each chopped pair was calculated and the nod frames were then differenced and combined to create a single image per HWP PA. During this process, all nods were examined for high and/or variable background that could indicate the presence of clouds or high precipitable water vapor, but no data needed to be removed for these reasons. Since Cygnus A was observed in two different sets of images, each HWP PA frame was registered and shifted to a common position, then images with the same HWP PA were co-averaged. Next, the ordinary (o-ray) and extraordinary (e-ray) rays, produced by the Wollaston prism,  were extracted and Stokes parameters, I, Q and U were estimated according to the ratio method prescription \cite[e.g.][]{Tinbergen2006}. Finally, the degree, $P = \sqrt{Q^2 + U^2}/I$, and PA, $PA = 0.5\arctan{(U/Q)}$, of polarization were estimated. 

The polarized Young Stellar Object, AFGL 2403, was observed in the 8.7 \um~filter immediately before Cygnus A. In total flux, AFGL 2403 was used as a point spread function (PSF) star, where the full width at half-maximum (FWHM) was 0.38$''$. A Moffat function with two parameters, FWHM and $\beta$ parameter, best describes the delivered PSF \cite[see][]{Radomski2008}. As seeing improves at longer wavelengths, and 0.38$''$ is much larger than the diffraction limit at 11.6 \um~($\sim$0.28$''$), this can be considered to be a conservative FWHM for the 11.6 \um~filter. In polarimetry, AFGL 2403 allowed us to estimate the zero-angle calibration of the observations. The zero-angle calibration, $\Delta\theta$, was estimated as the difference of the measured PA of polarization of our observations, $\theta = 54\pm8$\degree, and the PA of polarization measured by \citet{Smith2000}, $\theta_{s} = 40\pm8$\degree. Thus, the zero-angle calibration was estimated to be $\Delta\theta = \theta_{s} - \theta = -14$\degree. Our measured degree of polarization, $1.6\pm0.5\%$, of AFGL 2403 was in excellent agreement with the measured degree of polarization, $1.5\pm0.4\%$, by \citet{Smith2000}.  

The instrumental polarization was corrected based on data provided by the GTC website\footnote{Further information about the instrumental polarization of CanariCam is at http://www.gtc.iac.es/instruments/canaricam/\#Instrumental\_ Polarization}. Specifically, the instrumental polarization is P$_{\mbox{ins}}$ = 0.5$\pm$0.2\% in all filters with a dependency in the PA of polarization given by PA$_{\mbox{ins}}$ = -(RMA+Elev) - 29.6\degree, where RMA is the Nasmyth rotator mechanical angle; and Elev is the telescope elevation. The instrumental polarization was corrected as follow. The normalized Stokes parameters, q$_{\mbox{ins}}$ = Q$_{\mbox{ins}}$/I$_{\mbox{ins}}$ and u$_{\mbox{ins}}$ = U$_{\mbox{ins}}$/I$_{\mbox{ins}}$, of the instrumental polarization were estimated using the degree, P$_{\mbox{ins}}$, and position angle, PA$_{\mbox{ins}}$, of the instrumental polarization. Then, the q$_{\mbox{ins}}$ and u$_{\mbox{ins}}$ were subtracted from the normalized Stokes parameters of Cygnus A.

Dedicated flux standard stars were not observed. Flux calibration was performed using the 8-13 \um~spectrum acquired in a 0.5$''$ slit with the long wavelength spectrometer (LWS) on the 10.0-m {\it Keck I}~telescope by  \citet{IU2000}. The flux densities from the MIR spectra of Cygnus A are 50$\pm$10 mJy and 110$\pm$22 mJy at 8.7 \um~and 11.6 \um, respectively. Then, measured counts in a 0.5$''$ $\times$ 0.5$''$ simulated slit aperture from our images in the 8.7 \um~and 11.6 \um~filters were equated to the flux densities from the MIR spectra by \citet{IU2000}. Finally, the factor mJy/counts was estimated and used in the measurements of the flux densities in the several apertures presented in $\S$\ref{RES}.  As the dominant contributor to the uncertainty in the measurements of the total flux densities is based on the calibration from MIR spectroscopic data using {\it Keck I} by \citet{IU2000}, we used the uncertainty of 20\% for the measurements of the total flux densities in our observations.


\section{Results}
\label{RES}

We made measurements of the nuclear flux density in several apertures to compare with previously published values (Table \ref{Table1}). We measured the aperture flux density to be 29$\pm$6 mJy and  45$\pm$9 mJy in the 8.7 \um~and 11.6 \um~filters, respectively in an aperture diameter (hereafter aperture refers to diameter) equal to the FWHM, 0.38$''$ ($\sim$380 pc), of the PSF. This value optimally measures the nuclear flux density of Cygnus A and minimizes contamination from extended (diffuse) warm nuclear and emission from heated dust. The statistical significance of the aperture flux detections is $\sim$40$\sigma$ in both filters.


\begin{deluxetable}{cccllc}
\tablecaption{Aperture Flux Density and Polarization Measurements\label{Table1}}

\tablehead{
\colhead{Filter}	&	\colhead{Aperture}	&	\colhead{Flux}		&	\colhead{P}		&	\colhead{PA}				\\
				&	\colhead{diameter}	&	\colhead{density}	&					&							\\
				&	\colhead{($''$)}		&	\colhead{(mJy)}		&	\colhead{(\%)}	&	\colhead{(\degree)}	&			
}
\startdata
8.7 \um$^{a}$			&		0.38		&		29$\pm$6			&	11$\pm$3		&	27$\pm$8					\\
						&		2.0			&		101$\pm$20			&	\nodata			&	\nodata					\\
11.6 \um$^{a}$			&		0.38		&		45$\pm$9			&	12$\pm$3		&	35$\pm$8					\\
						&		2.0			&		169$\pm$34			&	\nodata			&	\nodata					\\
\hline
2.0 \um$^{b}$			&		0.375		&		4.9$\pm$0.1			&	10$\pm$1.5		&	201$\pm$3				\\	
N band$^{c}$ 			&		2.0			&		104$\pm$3			&	\nodata			&	\nodata							\\
11.7 \um$^{d}$			&		1.92		&		122$\pm$12			&	\nodata			&	\nodata					
\enddata

\tablerefs{(a) This work; (b) \citet{Tadhunter2000}; (c) \citet{Radomski2002}; (d) \citet{WA2004}}

\end{deluxetable}



\begin{figure*}[htbp]
\includegraphics[angle=0,trim=-5cm 0cm 0cm 0cm,scale=.30]{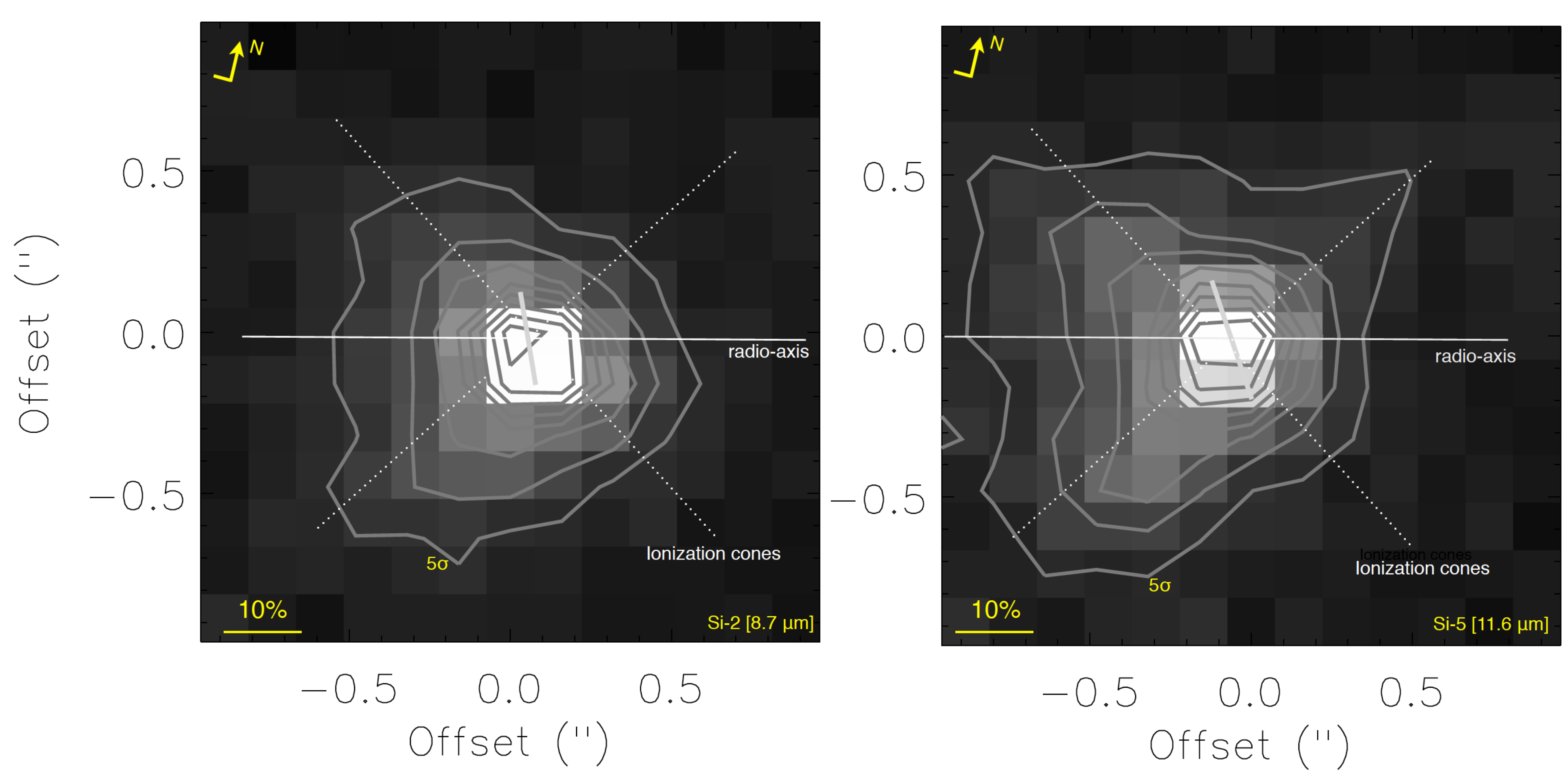}
\caption{Total flux ({\it grey-scale}) images and nuclear polarization vector in a 0.38$''$ ($\sim$380 pc) aperture of Cygnus A in the 8.7 \um~({\it left}) and 11.6 \um~({\it right}) filters. Contours start at 5$\sigma$ and increase in steps of 5$\sigma$. A polarization vector of 10\% polarization is shown ({\it bottom-left}). Total flux images were binned using a 2 $\times$ 2 pixels (0.16$''$$\times$0.16$''$) box only for display purposes. Radio-axis ({\it solid line}) with PA$_{jet}$ = 284$\pm$5\degree~\citep{Sorathia1996} and ionization cones ({\it dashed-line}; \citet{Tadhunter1999}) are shown.}
\label{fig1}
\end{figure*}


We measured the nuclear polarization using the I, Q and U images in a 0.38$''$ aperture, to match the FWHM of the observations. The degree of polarization was measured to be 11$\pm$3\% and 12$\pm$3\% with a PA of polarization measured to be 27$\pm$8\degree~and 35$\pm$8\degree in the 8.7 \um~and 11.6 \um~filters, respectively. The statistical significance of the polarization measurements is $\sim$4$\sigma$ of the polarization degree in both filters. The uncertainties in the degree of polarization and PA of polarization were estimated using the method of \citet{NC1993}, which estimates the uncertainties based on the SNR and measured degree and PA of polarization values.

Figure \ref{fig1} shows the total flux images in the 8.7 \um~and 11.6 \um~filters with the lowest contour at the 5$\sigma$ level. Due to the low SNR of the polarization detection, the polarization measurement is treated as a single statistically significant polarization vector at the 4$\sigma$ level of the polarization degree using the 0.38$''$ aperture in the 8.7 \um~and 11.6 \um~filters (Table \ref{Table1}). In total flux, the images in the 8.7 \um~and 11.6 \um~filters show a point source at the position of the nucleus with diffuse dust emission around the nuclear region with two extended emissions along the NE and SE arms of the ionization cones, previously observed in the 10.8 \um~broad-filter imaging~\citep{Radomski2002} and in the 11.7 \um~narrow-filter imaging \citep{WA2004}. In our 11.6 \um~filter observations, we detect diffuse extended dust emission in the NW arm at the 5$\sigma$ level in total flux. This diffuse extended dust emission component is spatially coincident with the extension previously observed in the 2.0 \um~filter using {\it HST}/NICMOS by \citet{Tadhunter1999}.

Previous NIR polarization measurements found a perpendicular PA of polarization with the radio jet in Cygnus A. Specifically, T00 measured a nuclear PA of polarization in the 2 \um~filter of PA$_{2\mu m} =$ 201$\pm$3\degree, almost perpendicular (83$\pm$6\degree) to the radio jet axis, PA$_{\mbox{jet}}$ = 284$\pm$5\degree~\citep{Sorathia1996}.We find the MIR PA of polarization is rotated by 77$\pm$13\degree~and 69$\pm$13\degree~in the 8.7 \um~and 11.6 \um~filters from the PA$_{\mbox{jet}}$.  In common with the 2.0 \um~observations, this is close to perpendicular to the PA$_{\mbox{jet}}$. That the estimation of the rotations are not exactly perpendicular to the radio jet axis can be attributed to (1) the relatively large uncertainty in the PA of polarization due to low SNR; and (2) the possible mix of several mechanisms of polarization ($\S$\ref{MODEL},\ref{DIS}) as well as different dominant polarization mechanisms at different wavelengths within the 0.38$''$ aperture.


\section{Polarization Model}
\label{MODEL}


\begin{figure*}[htpb]
\centering
\includegraphics[angle=0,trim=0cm 0cm 0cm 0cm,scale=.40]{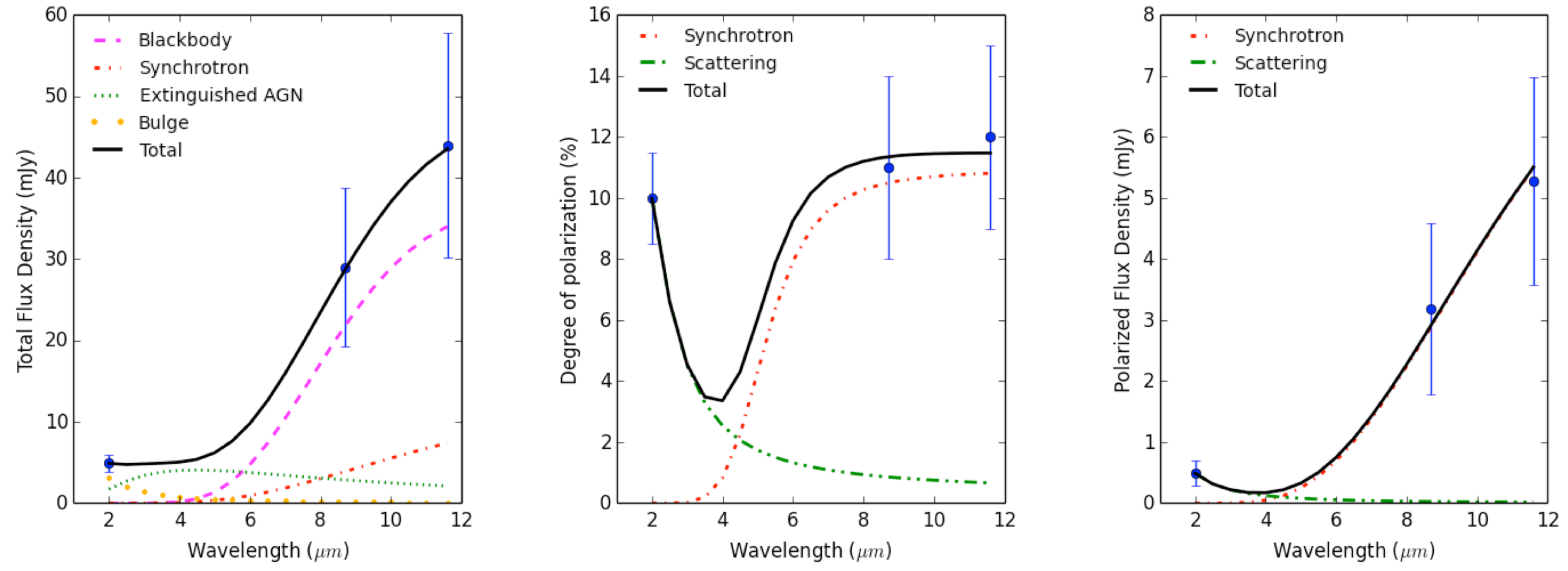}
\caption{Model fit to the measured total flux density ({\it left panel}) in a 0.38$''$ ($\sim$380 pc) aperture is shown. Measured total flux density ({\it blue dots}), nuclear bulge ({\it orange-circles}), extinguished AGN emission ({\it green-dots}), synchrotron radiation from the pc-scale jet ({\it red-dahed-dot}), blackbody emission at 220 K ({\it magenta-dashed}) and the total model ({\it black-solid}) are shown. Model fit to the measured degree of polarization ({\it middle panel}) and polarized flux density ({\it right panel}) in a 0.38$''$ ($\sim$380 pc) aperture are shown. Measured degree of polarization ({\it blue dots}), measured polarized flux density ({\it blue dots}), synchrotron radiation from the pc-scale jet ({\it red-dashed-dot-dot}), dust scattering ({\it green-dashed-dot}) and total model ({\it black-solid}) are shown.}
\label{fig2}
\end{figure*}


We aim to reproduce the observed polarization from 2.0 \um~to 11.6 \um. Specifically, we need a polarization component that reproduces the observed polarized flux rising with increasing wavelength. We developed a polarization model that simultaneously fits the total flux, degree of polarization, and polarized flux. In the model, we consider the following components: (1) stellar emission from the nuclear bulge; (2) extinguished AGN emission; (3) scattered radiation by dust within the unresolved nuclear region; (4) thermal emission from dust (torus and/or diffuse dust emission);  and (5) synchrotron radiation from the pc-scale jet. To constrain the polarization mechanisms, we used our polarimetric measurements in a 0.38$''$ aperture in the 8.7 \um~and 11.6 \um~filters (Table \ref{Table1}), in combination with the measured degree of polarization in the 2.0 \um~filter of 10$\pm$1.5\%  by T00. This measurement has similar aperture size to our MIR measurements and minimizes contamination from extended components and starlight. 

Starlight emission and its possible polarization are negligible at 8.7 \um~and 11.6 \um. T00 estimated the contribution of the AGN emission to be 35\% at 2.0 \um~through PSF-subtraction and, we assume that the stellar component is 65\% of the total flux density. We use the 2.0 \um~flux density data to estimate the flux density in our MIR filters, using a simple and approximate color correction of the total flux density in K (2.2 \um)~to~N-band (10 \um), S$_{\mbox{N}}$ = 0.091S$_{\mbox{K}}$ \citep{KGW1992}. This yields a flux density of 0.3 mJy in our 0.38$''$ aperture in the N band, which is $\sim$1\% of the aperture flux density at 8.7 \um, and a dichroic polarization of 0.1\% \citep{Serkowski1975}. Both, aperture flux density and dichroic polarization, are lower at 11.6 \um. Thus, the dichroic absorption of starlight is considered to be negligible given the high measured degree of polarization of Cygnus A within the 2-11.6 \um~wavelength range.

The AGN emission, defined to be the emission from the accretion disk of the central engine, is assumed to be an extinguished unpolarized power-law, F$_{\lambda} \propto \lambda^{-\alpha_{AGN}} e^{-\tau_{\lambda}}$, where $\tau_{\lambda}$, is the absorption produced by the dust screen (i.e. dust lane). We adopted a spectral-index $\alpha_{AGN}$ = 1.8 from the intrinsic infrared AGN SED template in the wavelength range of 6-19 \um~\citep{Mullaney2011}.  As in the case of the bulge component, this simple AGN power-law ensures no additional free parameters, silicate features and/or uncertainties in the polarization model. Recent studies have demonstrated that a ``universal'' NIR to MIR extinction law cannot be defined \citep[e.g.][]{FM2009,GJA2009}. However, standard silicate-graphite interstellar grain for R$_{\mbox{v}}$ = 5.5 appears to be in agreement with the MIR extinction curves at different sightlines through the Galactic Center \citep{WD2001}. We took the extinction law\footnote{Note that recent studies \citep{GLJ2013} suggested that a trimodal grain size distribution with a combination of R$_{\mbox{v}}$ = 2.1, 3.1 and 5.5 is required to achieve a good fit to the extinction curve from the UV to the MIR.} for R$_{\mbox{v}}$ = 5.5, with A$_{\mbox{8.7\um}}$  $\sim$ A$_{\mbox{11.6\um}}$ = 0.75 A$_{\mbox{K}}$ \citep[fig. 1 in][]{GLJ2013}, and A$_{\mbox{K}} =$ 0.112 A$_{\mbox{V}}$ \citep{Jones1989}.  We took the visual extinction of A$_{\mbox{v}}$ = 94 mag to the nucleus of Cygnus A \citep{Tadhunter1999}, because it represents the most comparable value to our aperture size and wavelength.

As noted in the introduction, dust scattering is the dominant polarization mechanism in the central regions of Cygnus A at optical wavelengths \citep{Tadhunter1990,JT1993,Ogle1997,vanBemmel2003}. To study how dust scattering affects polarization in the unresolved nucleus of Cygnus A within the 2-11.6 \um~wavelength range, this mechanism is included in our polarization model. We assumed a wavelength dependency in the total flux density to be F$^{sca} \propto \lambda^{-4}$, consistent with scattered radiation by dust within the unresolved core. The degree of polarization is also a function of the wavelength, P$^{sca} \propto \lambda^{-4}$. We took the intrinsic degree of polarization of 28\% estimated by T00 at 2.0 \um, to estimate the observed polarization in our polarization model.

For MIR thermal emission in the core of Cygnus A, \citet{Radomski2002} estimated that temperature reaches a value of 150$\pm$10 K up to 500 pc from the central source, with a lower-limit of 220$\pm$30 K in the NE cone and 150$\pm$5 K in the SE cone. \citet{WA2004} estimated a temperature of 120 K up to 500 pc from the core of Cygnus A. \citet{Merlo2014} suggested that the MIR emission is produced by blackbody component with a characteristic temperature of 217$\pm$3 K in our LOS to the core. Based on these studies and the physical size ($\sim$380 pc) of our aperture, the final characteristic temperature of the blackbody component in our polarization model was estimated as the best fit of a blackbody component in steps of 10 K within the range of 120-220 K. We assume this blackbody component arises from the diffuse dust emission around the nuclear region and/or dusty torus. This component is assumed to be unpolarized and no dichroism is considered in the polarization model ($\S$\ref{DIS}).

As noted in the introduction, previous studies have observed pc-scaled jets that contribute, at some fraction, to the total flux density as a synchrotron component with a cut-off wavelength in the MIR wavelengths. In our model, synchrotron emission is described as a cut-off wavelength power-law source behind a dust screen (i.e. dust lane), F$_{\nu}^{syn} \propto e^{-\nu/\nu_{c}} \nu^{-\alpha_{syn}} e^{-\tau_{\lambda}}$, where $\nu_{c}$, is the cut-off frequency; $\alpha_{syn}$, is the spectral index; and $\tau_{\lambda}$, is the absorption produced by the dust screen. We took the spectral-index, $\alpha_{syn}$, of 0.18 estimated using radio {\it VLBA} observations and previously used in the SED models by P12. Recent polarimetric studies \citep{Ramirez2014} in the 2.0 \um~filter using {\it HST}/NICMOS, in combination with radio and millimeter wavelengths, of a sample of narrow-line radio galaxies (Cygnus A included) estimated an averaged spectral index of 0.21$\pm$0.1 (0.18 for Cygnus A). The pc-scale jet has been resolved using {\it VLBA} observations \citep{Sorathia1996}, hence the only extinction along our LOS to be considered is that produced by the dust lane of the Cygnus A galaxy. We took the visual extinction of the dust lane of 5.5 mag as modeled by \citet{Packham1998}. The polarized flux density of the synchrotron radiation follows the same power-law source that the total flux density. 
 
With the constraints on each component given above, we simultaneously fit the measured total flux density, degree of polarization and, polarized flux (Figure \ref{fig2}). The fit was considered acceptable when the deviations from the modeled total and polarized flux values were $<$5\% of the measured values at 2.0 \um, 8.7 \um~and 11.6 \um. For the best fit, we found a blackbody component with a characteristic temperature of 220 K and a synchrotron cut-off wavelength at 34 \um. Within an aperture of 0.38$"$ ($\sim$380 pc), the fraction of total and polarized flux density for each one of the components in our polarization model are shown in Table \ref{Table2}. Note that this polarization model is excluding polarization from dichroic emission, which can contribute at a low level to the total polarized flux density.

\begin{deluxetable}{lccc}
\tablecaption{Fraction of aperture and polarized flux density for each mechanism in the central 380 pc of Cygnus A\label{Table2}}

\tablehead{
\colhead{Component}	&	\colhead{2.0 \um}	&	\colhead{8.7 \um}	&	\colhead{11.6 \um}
}
\startdata
							&	\multicolumn{3}{c}{Total Flux density} \\
\hline
Nuclear bulge 			&		65\%			&		negligible			&		negligible		\\
Extinguished AGN		&		35\%			&		10\%				&		5\%			\\
Synchrotron				&		negligible		&		14\%				&		17\%			\\
Blackbody radiation		&		negligible		&		75\%				&		78\%			\\
\hline
					&	\multicolumn{3}{c}{Polarized Flux density} \\
\hline
Dust scattering 		&		$>$99\%			&		negligible			&		negligible		\\
Synchrotron			&		negligible			&		$>$99\%			&		$>$99\%			
\enddata

\end{deluxetable}


\section{Discussion}
\label{DIS}

The most plausible explanation of the polarization in the nucleus of Cygnus A is that the MIR polarization dominantly arises from synchrotron radiation. The intrinsically polarized source flux is diluted by the blackbody component with characteristic temperature of 220 K. The intrinsic polarization is estimated to be  P$^{int}_{syn} \sim$ 65\% in the MIR wavelengths. This value is in agreement with a linear polarization as high as 70\% produced by optically thin synchrotron radiation. The high intrinsic degree of polarization implies a very ordered magnetic field within the pc-scale jet close to the core of Cygnus A. This characteristic has been previously observed in Cygnus A \citep{Krichbaum1998} and other AGN, e.g. NGC 1052 \citep{Vermeulen2003}. Also, shock waves within the jets compress the magnetic field and create regions of ordered magnetic fields \citep{Laing1980}.  A high intrinsic polarization with the PA of polarization approximately perpendicular to the pc-scale jet are readily produced by synchrotron radiation from the pc-scale jet observed at radio wavelengths \citep{Krichbaum1998}. The PA of polarization is coincident for both the NIR and MIR wavelengths.  The PA of polarization arising from scattering within the innermost region of the central engine of Cygnus A will be aligned with the major axis of the torus, i.e. perpendicular to the polar axis. This implies that the PA of polarization in the NIR and MIR will be nearly the same, but that the dominant mechanisms of polarization are not the same.

The polarization model explains the observed MIR total flux density from the contribution of (1) a blackbody component with a characteristic temperature of 220 K; and (2) a synchrotron component. The latter only contributes 14\% and 17\% of the observed nuclear total flux density in the 8.7 \um~and 11.6 \um~filters, respectively (Table \ref{Table2}). From our polarization model, the blackbody emission with a characteristic temperature of 220 K can arise from (a) dust in the torus; and/or (b) diffuse dust emission component in our LOS. The temperature is consistent with directly radiated dust from the central engine located at a radius of $\sim$130 pc, as well as, emitted radiation from dust in the non-directly illuminated faces of the clumps within the torus at scales of few parsecs from the central engine. Thus, these components cannot be distinguished from the current model. 

Several NIR polarimetric studies \citep[e.g.][]{JK1989,Y1995,Packham1998,Simpson2002} of AGN have found that some of the nuclear polarization of type 2 AGN often arises from the passage of unpolarized radiation from the central engine through aligned dust grains in the torus in our LOS. To investigate the effect this mechanism could have on our observations, we investigate this mechanism here. \citet{JKD1992} explained the trend of polarization with optical depth in the interstellar medium using grains aligned with a magnetic field that was a mixture of constant and random components. If the visual extinction of 94 mag \citep{Tadhunter1999} to the nucleus of Cygnus A  is assumed, the predicted level of polarization at K produced by dichroic absorption is 51\% for constant component of the magnetic field, 3\% for random alignment of the magnetic field, and 11\% for equal contribution of the constant and random component of the magnetic field. If we assume that the intrinsic polarization in the 2.0 \um~filter of 28\% (T00) arises from dichroic absorption of unpolarized radiation of the central engine by aligned dust grains in our LOS, then Cygnus A has a constant component of the magnetic field dominating the thermal motions in determining the alignment geometry of dust grains. Although some dust grain alignment (although far from perfect alignment) is expected, it is difficult to distinguish where the dichroic absorption arises: (1) within the torus; or (2) from the dusty blackbody component in our LOS. We believe that, in agreement with T00, such a high, 28\%, intrinsic degree of polarization at 2.0 \um~is far more likely to arise from a scattering region close to the central engine of Cygnus A rather than dichroism. Further, we note that such a high level of intrinsic of polarization produced by dichroism is rarely observed in other AGN at NIR wavelengths \citep[e.g.][]{LR2013,Ramirez2014}. Thus, although we cannot exclude the possibility of a dichroic absorption source of the polarization, we find it very unlikely. 

\citet{Veret2001} showed that the UV-optical continuum of several quasars at z$\sim$2.5 can be explained by dust-reflected quasar light resulting in grey scattering (large dust grains produce scattering with a constant degree of polarization) from a highly clumped scattering medium close to the nucleus. We studied the possibility of large grains in a clumped medium close to the core of Cygnus A. This clumped medium produces a grey scattering component, reddened by the blackbody component in our LOS, which can explain the polarized flux density rising with increasing wavelength. Large grains are considered when grain sizes are the order of the incident radiation wavelength. In this case, grain sizes of $\sim$1 \um~and $\sim$10 \um~from NIR and MIR radiation, respectively are required. These grains sizes are very difficult to conceive with typical ISM dust grains, where typical grain models \citep{MRN1977} suggest grain sizes between 0.005-0.25 \um. Although the observed polarized flux density rising with wavelength in Cygnus A (Figure \ref{fig2}) can be fit by a grey scattering model, this mechanism is difficult to conceive given the large grain sizes that are required to fit the data.

\section{Conclusions}

We found that synchrotron radiation from a pc-scale jet close to the core of Cygnus A is the most likely mechanism to explain the  polarized flux rising with increasing wavelength. Based on our developed polarization model, the synchrotron radiation from the pc-scale jet is estimated to be 14\% and 17\% of the total flux density in the 8.7 \um~and 11.6 \um~filters, respectively. A blackbody component with a characteristic temperature of 220 K can account for  $>75\%$ of the observed MIR total flux density.  The blackbody emission can arise from (1) dust in the torus; and/or (2) diffuse dust emission component in our LOS, but these two components cannot be distinguished from the current observations. These observations represent the most compelling detection of a synchrotron component using MIR polarimetric observations in Cygnus A. Future similar observations of a sample of radio galaxies will be interesting to determine general and/or extraordinary properties in these objects.




\acknowledgments

We would like to thank the anonymous referee for his or her useful comments, which improved the paper significantly. Based on observations made with the {\it Gran Telescopio CANARIAS} ({\it GTC}), installed in the Spanish Observatorio del Roque de los Muchachos of the Instituto de Astrof\'isica de Canarias, in the island of La Palma. ELR acknowledges support from an University of Florida Alumni Fellowship to make possible this work as part of his thesis project and also acknowledges the financial support of University of Texas at San Antonio. CP acknowledges support from NSF-0904421 grant. AAH acknowledges support from the Spanish Plan Nacional through grant AYA2009-05705-E. CRA ackowledges support from the Marie Curie Intra European Fellowship PIEF-GA-2012-327934 within the 7th European Community Framework Programme and from the Spanish Plan Nacional through grant AYA2010-21887-C04.04 (Estallidos). RM and NAL are supported by the Gemini Observatory, which is operated by the Association of  Universities for Research in Astronomy, Inc., on behalf of the international  Gemini partnership of Argentina, Australia, Brazil, Canada, Chile, and the United States of America. EP acknowledges support from AST-0904896.

\clearpage

\end{document}